# INSTRUMENTAL INTENSITY AS A TOOL FOR POST-EARTHQUAKE DAMAGE ASSESSMENT: VALIDATION FOR THE STRONG VRANCEA EARTHQUAKES OF AUGUST 1986 AND MAY 1990

*Iolanda-Gabriela CRAIFALEANU* [1], *Ioan Sorin BORCIA* [2]

**ABSTRACT**

The frequency-dependent spectrum based seismic intensity, also called instrumental intensity, is calculated basically from the integration of the square values of spectral acceleration ordinates. The values of the instrumental intensity are calibrated to match the values of the EMS-98 intensity scale, providing a promising analytical indicator for estimating the destructive potential of earthquakes. Previous studies have shown that the proposed index could be used as a basis for the development of a new improved seismic intensity scale. The paper presents a set of maps describing the spatial distribution of instrumental intensity ordinates for three seismic events recorded in 1986 and 1990. These events, generated by the Vrancea source, are the strongest earthquakes in Romania for which accelerographic data was recorded at multiple stations. Intensity maps were generated for separate significant frequency bands, in order to reveal the destructiveness of the considered earthquakes for different building categories. Results were compared and correlated with previous studies on Vrancea earthquakes and with information provided by building damage reports from the considered earthquakes.

**REZUMAT**

Intensitatea seismică bazată pe spectrul de răspuns, denumită şi intensitate instrumentală, este calculată pe baza integrării pătratelor valorilor acceleraţiilor spectrale. Valorile intensităţii instrumentale sunt calibrate astfel încât să corespundă valorilor scării de intensitate seismică EMS-98, reprezentând un indicator analitic promiţător în estimarea potenţialului distructiv al seismelor. Studii anterioare au arătat că intensitatea seismică instrumentală ar putea fi utilizată drept bază pentru dezvoltarea unei scări de intensitate seismică îmbunătăţite. Articolul prezintă o serie de hărţi care descriu distribuţia spaţială a ordonatelor intensităţii instrumentale pentru 3 evenimente seismice înregistrate în 1986 şi 1990. Aceste evenimente, generate de sursa Vrancea, sunt cele mai puternice cutremure produse în România, pentru care au fost înregistrate date accelerografice în multiple staţii. Au fost generate hărţi de intensitate seismică pentru benzi de frecvenţe considerate semnificative, cu scopul de a releva potenţialul distructiv al cutremurelor menţionate, pentru diferite categorii de clădiri. Rezultatele au fost comparate şi corelate cu studiile existente referitoare la cutremurele vrâncene şi cu informaţiile furnizate de rapoartele privind avariile produse clădirilor de seismele respective.

*Cuvinte cheie:* intensitatea seismică bazată pe spectru, cutremure vrâncene, scara de intensitate seismică EMS-98, hărţi de intensitate seismică

[1] Corresponding author: Technical University of Civil Engineering Bucharest and National Institute for Research and Development in Construction, Urban Planning and Sustainable Spatial Development "URBAN-INCERC", INCERC Bucharest Branch; e-mail: i.craifaleanu@gmail.com, iolanda@incerc2004.ro

[2] National Research and Development Institute URBAN-INCERC, INCERC Bucharest Branch; e-mail: isborcia@incerc2004.ro

## 1. INTRODUCTION

The intensity based on destructiveness spectrum, $i_d(\varphi)$ is defined by the following expression [Sandi, 1987, Sandi et al., 1998]

$$i_d(\varphi) = \log_4 \left( \int w_a^2(t,\varphi,\xi) \, dt \right) + 5.75 \quad (1)$$

where $w_a(t,\varphi,\xi)$ is the absolute acceleration (m / s²), for a pendulum of natural frequency $\varphi$ (Hz), and $\xi = 5\%$ is the damping ratio. The values of the above instrumental intensity are calibrated to match the values of the EMS-98 intensity scale.

In order to assess the destructiveness on separate frequency bands, the intensity in equation (1) was averaged upon spectral bands, $(\varphi', \varphi'')$, the averaging rule being described by the following equation:

$$i_d^*(\varphi',\varphi'') = \log_{7.5}\{1/\ln(\varphi''/\varphi') \cdot$$
$$\cdot \int [(\int w_a^2(t,\varphi,\xi) \, dt) \, d\varphi/\varphi]\} + 6.45 \quad (2)$$

The analysis was related to the 36 dB frequency band (0.25 Hz…16.0 Hz), adopted as a reference. This was divided into twelve 3 dB subintervals. The Id12 intensity values calculated for these 12 subintervals were denoted, in order, by Id121, Id122 … Id1212. From these intervals, only 4, considered as characteristic for a large part of the frequency range of building structures, were studied, as shown in Table 1. For convenience, frequency intervals were expressed as period intervals. Results obtained by using different other frequency intervals were presented in [Craifaleanu and Borcia, 2010 and 2011].

**Table 1.**

*Symbols used to denote averaged instrumental intensities and corresponding intervals used for 12-subinterval averaging*

| Symbol | T″(s) | T′(s) |
|---|---|---|
| Id 124 | 1.00 | 1.41 |
| Id 125 | 0.71 | 1.00 |
| Id 126 | 0.50 | 0.71 |
| Id 127 | 0.35 | 0.50 |

## 2. INSTRUMENTAL INTENSITY MAPS

The maps of the spatial distribution of intensities in Table 1 were generated, for the three strongest Vrancea earthquakes for which accelerographic data was recorded at multiple stations. These are the earthquakes of August 30, 1986 ($M_w$ = 7.1, focal depth $h$ = 133 km), May 30, 1990 ($M_w$ = 6.9, $h$ = 91 km), and May 31, 1990 ($M_w$ = 6.4, $h$ = 87 km).

It should be noted that the set of stations providing seismic data was different for each seismic event, either due to the gradual expansion of seismic networks in Romania, or to the accidental malfunction of some instruments. Furthermore, the uneven spatial distribution of stations did not allow, especially for intra-Carpathian areas, obtaining reliable contours. On the other hand, the number of stations in the extra-Carpathian area and, particularly, in the epicentral area and in Bucharest, allowed for a quite thorough study of these especially exposed areas (Craifaleanu and Borcia, 2010).

The resulting maps are presented in the following. In order to facilitate interpretation, maps of Id12 intensities were arranged in order of increasing period.

An important observation that concerns all the maps is that the contour orientation, for each seismic event, is in quite good agreement with previous directivity studies (Sandi et al., 2004). According to the cited reference, the radiation directivity was radically different for the three events considered, i.e. approximately NE-SW on August 30, 1986, N-S on May 30, 1990 and S-E on May 31, 1990. This directivity can also be discerned on the seismic intensity maps, even though with less clarity, due to the fact that only the maximum value from the two components was considered in generating the maps.

Some observations concerning the maps generated for each seismic event are presented in the following.

a) Observations concerning the August 30, 1986 earthquake (Fig. 1).

1. The largest Id12 values (Id12 > 7) occur, for all frequency intervals, in a well-delimited area, oriented approximately along the NE-SW line connecting stations Peris (PRS1) and Chisinau (CHS1). The shape of this area is in good agreement with the macroseismic intensity maps compiled after the event (Radu et al., 1986).

2. In what concerns the dependence of instrumental intensity on frequency (period), the largest values (over 7.5) occur, generally, for $T = 0.35…0.5$ s (Id127), with values of over 8 in several stations located along the NE-SW line. These values tend to decrease rather monotonically with period, so that the lowest intensity values occur for the period range $T = 1…1.41$ s. However, even for this last period range, the instrumental intensity values still remain near to or greater than 7 in several stations, with largest values of Id124 = 8.13 at Focsani (FOC3), Id124 = 7.88 at Chisinau (CHS1) and Id124 = 7.86 at Valenii de Munte (VLM1).

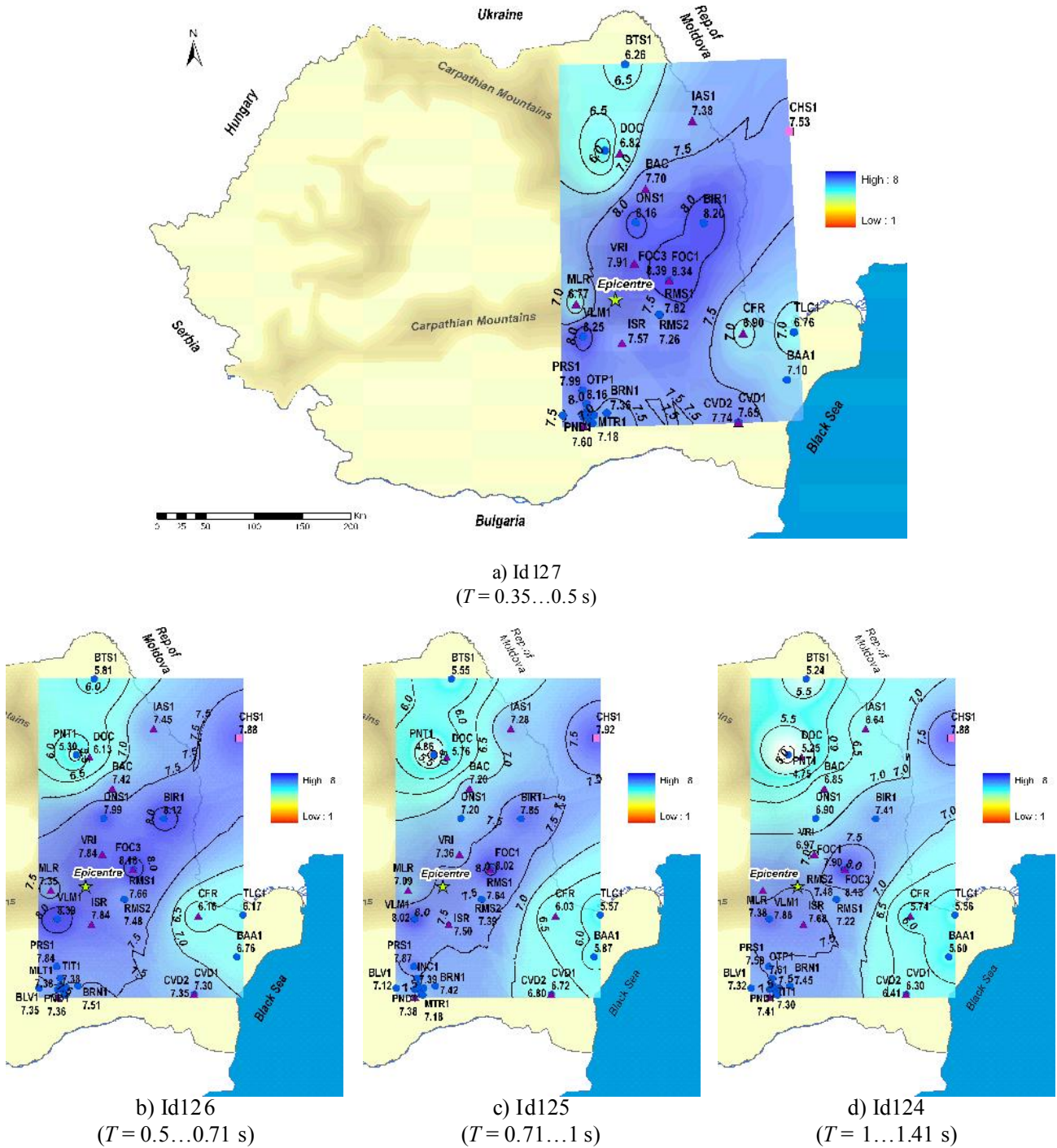

a) Id127
($T = 0.35…0.5$ s)

b) Id126
($T = 0.5…0.71$ s)

c) Id125
($T = 0.71…1$ s)

d) Id124
($T = 1…1.41$ s)

**Fig. 1.** Spatial distribution of Id127, Id126, Id125 and Id124 instrumental intensities, for the August 30, 1986 earthquake

b) Observations concerning the May 30, 1990 earthquake (Fig. 2).

1. The map contour pattern varies significantly with period, due to the differences between the types of variation recorded in each station. While the general decreasing tendency of the intensity values with period - observable in 1986 at most of the stations - is present here as well, there are some stations at which, during the specific frequency contents, the tendency is opposite, or the variation of the intensity values with period is non-monotonic. This is due to the differences in the frequency content of the ground motions, which influence the shape of acceleration spectra (see the definition of the seismic intensity in Equation 1). It must be, however, mentioned that these irregular variations of the seismic intensity are of small amplitude, as, for instance, at the Ramnicu Sarat stations (RMS1 and RMS2) located north from epicentre, and at the Calarasi (CLS1) and Baia (BAA1) stations, situated in the south-eastern part of the country.

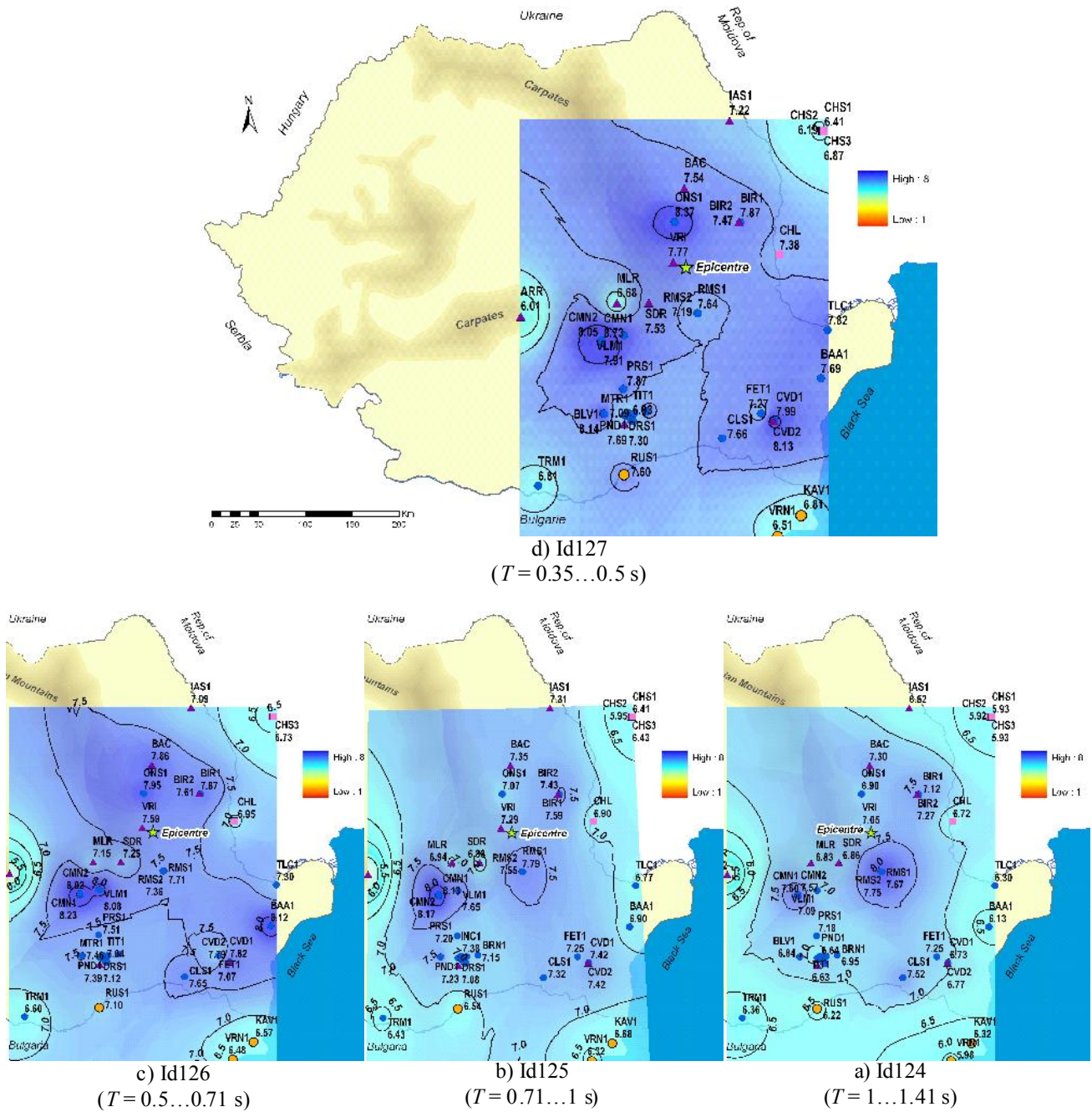

d) Id127
($T = 0.35…0.5$ s)

c) Id126
($T = 0.5…0.71$ s)

b) Id125
($T = 0.71…1$ s)

a) Id124
($T = 1…1.41$ s)

**Fig. 2.** Spatial distribution of Id127, Id126, Id125 and Id124 instrumental intensities, for the May 30, 1990 earthquake

2. For the short period range ($T = 0.35…0.5$ s), the largest seismic intensity values (Id12 > 8) occur at stations Campina (CMN1), Id127 = 8.73, Onesti (ONS1), Id127 = 8.37, Bolintin Vale (BLV1, north from Bucharest), Id127 = 8.14, and Cernavoda (CVD1, in the south-eastern part of the country), Id127 = 8.13. For the second period range considered ($T = 0.5…0.71$ s), the seismic intensity contour pattern is quite similar, with a remarkable value of Id126 = 8.12 at station Baia (BAA1).

3. For the third and fourth period ranges considered ($T = 0.71…1$ s and $T = 1…1.41$ s), the largest seismic intensity values occur at stations Campina (CMN1 and CMN2), Ramnicu Sarat (RMS1 and RMS2) and Barlad (BIR1 and BIR2).

4. It is worth noting that stations as Chisinau (CHS), the stations located on the Bulgarian seashore (Kavarna, KAV1, and Varna, VRN1) or the station Turnu Magurele (TRM1), located in the southwest of the studied zone, have systematically smaller values than the stations situated in the predominant area of propagation of the seismic waves.

c) Observations concerning the May 31, 1990 earthquake (Fig. 3).

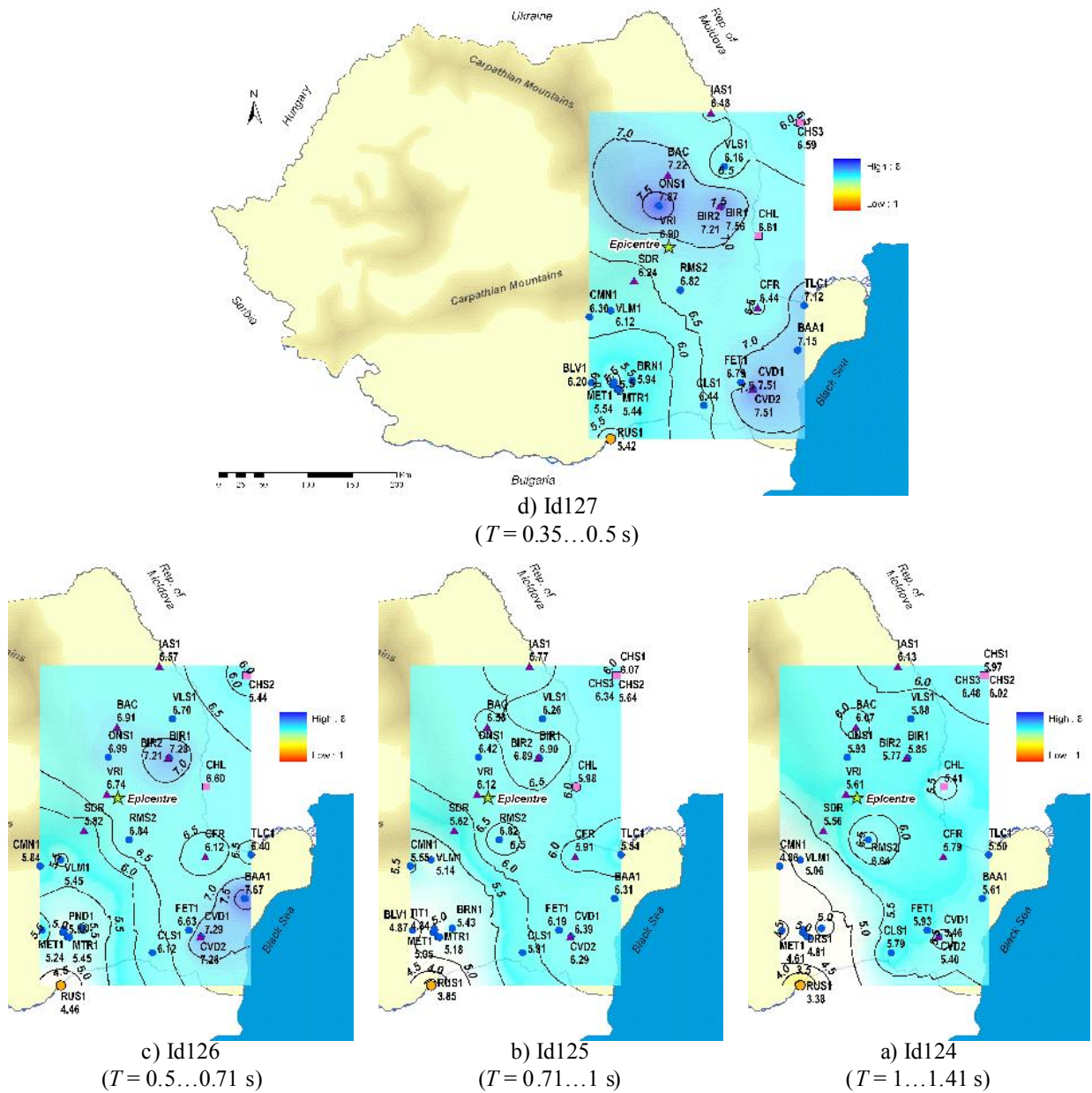

d) Id127 ($T = 0.35…0.5$ s)

c) Id126 ($T = 0.5…0.71$ s)

b) Id125 ($T = 0.71…1$ s)

a) Id124 ($T = 1…1.41$ s)

**Fig. 3.** Spatial distribution of Id127, Id126, Id125 and Id124 instrumental intensities, for the May 31, 1990 earthquake

1. For short periods (T = 0.35…0.5 s), the largest Id12 values (Id12 > 7) occur in the north-eastern part of the country (Onesti (ONS1), Barlad (BIR1), Bacau (BAC) and Vrancioaia (VRI) stations), and also in the south-eastern part of the country, at stations Cernavoda (CVD), Baia (BAA1) and Tulcea (TLC1). As in the case of the August 31, 1986 and May 30, 1990 earthquakes, seismic intensity values tend to decrease, generally, with period, with some exceptions as, for instance, the Chisinau station (CHS).

2. For the analyzed earthquake, the shape of the resulted intensity contours differs substantially from the one on the macroseismic intensity maps compiled after the event (Radu and Utale, 1991), where the contour for values greater than 6 has a similar shape with that for the May 30, 1990 event, and where values over 7 are assigned in a small area located at the north-east of the epicentre. These differences can be put down to the peculiarities of the assessment of macroseismic intensity.

## 3. DAMAGE REPORTED AT BUILDINGS FOLLOWING THE 1986 AND 1990 EARTHQUAKES

Data on the effects of the of the August 30, 1986 and May 30, 1990 in Romania and in the Republic of Moldova were collected from various reports and publications (MTCT, 2002, Alcaz, 2006, Crainic et al., 1992, Zaicenco et al., 2004, USGS etc.).

The August 30, 1986 earthquake caused moderate losses in Romania. The damage ranged from non-structural to moderate structural in various types of buildings. Significant damage to some historical buildings, as well as damage or even collapse of church towers was also reported, in Focsani and in other cities. Additional damage occurred in buildings affected by the March 4, 1977 and which were not strengthened after this event. The rigid non-structural components of several buildings were often heavily affected (Crainic et al., 1992). In the Republic of Moldova, the earthquake caused much heavier losses (Zaicenco et al., 2004,

Alcaz, 2006). Several buildings suffered extensive damage in Kishinev and Cahul, as well as in other towns and villages of the country. Two people were killed and 561 injured, over 14000 people were left homeless. According to the Institute for Building Economics of the USSR, direct losses produced in the Republic of Moldova by the earthquake were evaluated to about 680 million USD.

The May 30, 1990 earthquake caused only minor structural damage to a small number of buildings in Romania, especially to those weakened by previous earthquakes. However, non-structural damage was present in many buildings. For instance, falling of façade cladding at the seismic gap between two buildings caused the death of two people in Bucharest. Seven other people died from similar causes and many others were injured. No structural damage to buildings in Republic of Moldova (Zaicenco et al., 2004) or in north-eastern Bulgaria was observed due to this earthquake.

## 4. CONCLUSIONS

The paper presented the first maps describing the spatial distribution of the instrumental intensity, for three strong Vrancea earthquakes, i.e. the August 30, 1986, May 30, 1990 and May 31, 1990 events.

This intensity index, proposed by H. Sandi in 1987, has, as compared with macroseismic intensity, the advantage of being based on instrumental data. The values of the instrumental intensity are calibrated to match the values of the EMS-98 intensity scale. Previous studies have shown that the proposed index could be used as a basis for the development of a new intensity scale, which would rely on instrumental data.

The maps confirmed the capacity of the instrumental intensity to express the destructive potential of a seismic event, by showing very good agreement with the macroseismic intensity maps. Moreover, the shape and predominant orientation of map contours matched the results of previous directivity studies performed for the concerned Vrancea events.

An important remark concerning all analyzed seismic events is that, at most stations, the largest values of the averaged instrumental intensity occurred for the first period range $T = 0.35…0.5$ s, which gives an indication of the characteristics (fundamental periods of vibration) of the potentially most affected building stock. Damage reports compiled after the 1986 and 1990 earthquakes are consistent with this indication. However, from the quantitative point of view, the estimations based on seismic intensity appear to be more severe than the effects actually reported as a consequence of the analyzed earthquakes.

**ACKNOWLEDGEMENTS**